# Sensor Insoles: A Review

Bastian Latsch, Felix Herbst, Mark Suppelt, Julian Seiler, Stephan Schaumann, Sven Suppelt, Alexander A. Altmann, Martin Grimmer, and Mario Kupnik, *Senior Member, IEEE*

***Abstract*—Plantar pressure measurement, or pedobarography, is an essential tool for analyzing human motion in healthy individuals and patients. Across the reviewed literature, sensor insoles are motivated as wearable, mobile solutions for assessing pressure distribution in applications including diabetic foot monitoring, rehabilitation guidance, assistive device control, and sports performance analysis. This review evaluates the current state of the art with particular attention to sensor technologies, sensor quantity and placement, participant cohorts, and reference standards. The focus lies on original works with innovative designs, preferably supported by ambulation experiments. The modalities covered include resistive, capacitive, inductive, piezoelectric, triboelectric, and optical sensing approaches. We identify a lack of proper sensor calibration, gait-based verification, and human study validation, and propose a gold standard based on testing machines and instrumented treadmills to ensure comparability across studies. The bidirectional interaction between insole insertion and foot-sole mechanics is examined, with tissue stiffness identified as a key source of uncertainty in sensor signals. Guidelines are provided for sensor dimensions and unobtrusive insole designs to foster natural gait. Finally, future directions include the development of multimodal sensors to compensate for the limitations of individual modalities and the emerging trend of multiaxial sensing for capturing shear components in pressure distributions.**

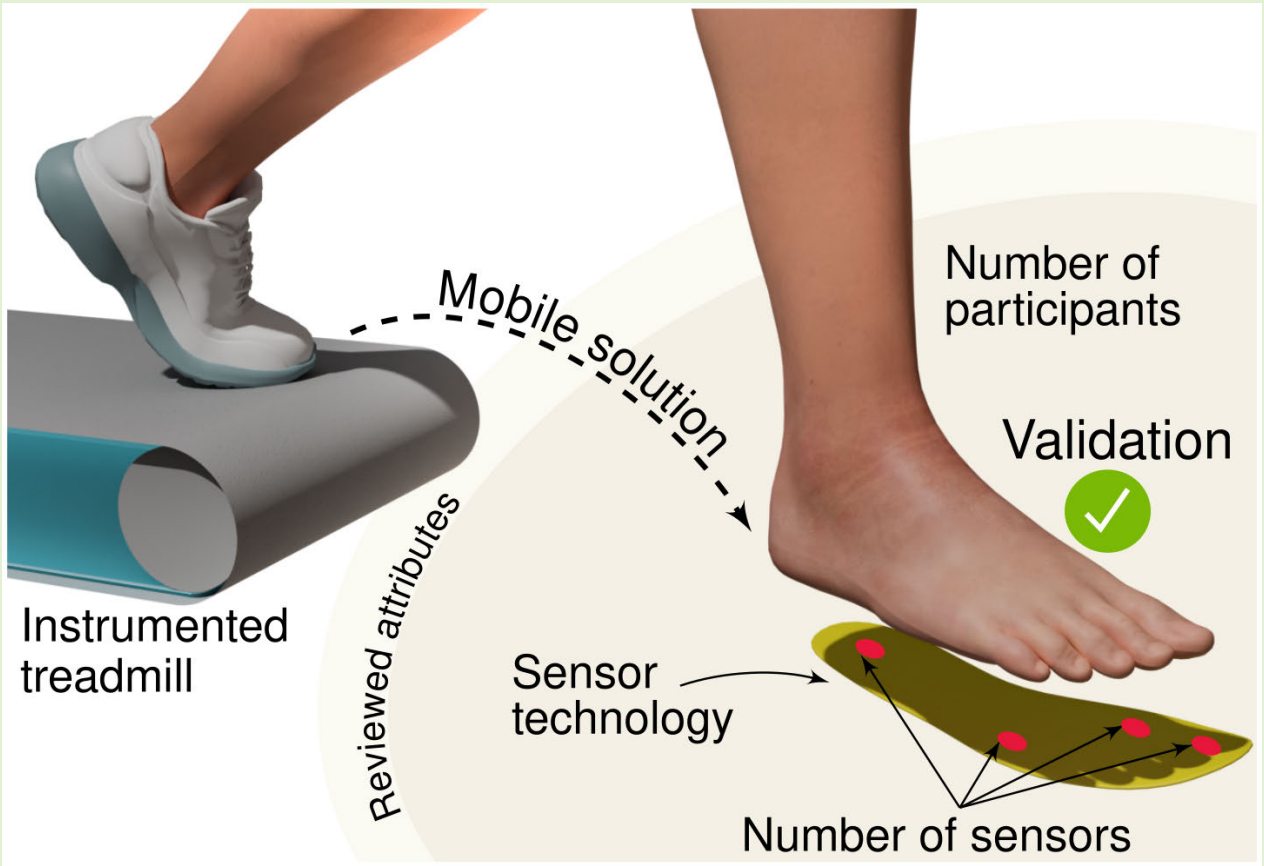

## I. Introduction

HUMAN locomotion, as a core component of physical activity, is a fundamental aspect of personal autonomy and social participation, significantly influencing independence and overall quality of life [1], [2]. Bipedal locomotion depends on a complex coordination of motor and cognitive functions, even to just maintain an upright posture. The underlying biomechanical processes are governed by a multitude of sensory inputs and vary from person to person. A variety of disturbances to this finely tuned system can significantly affect its stability, including gait-related impairments due to congenital deficiencies, disorders, age-related degeneration, injuries, and diseases [3], [4], [5], [6].

Analyzing human motion is an essential part of the anamnesis in impaired patients. It is used to evaluate the demand for assistive devices and to track rehabilitation progress. Human motion analysis primarily focuses on observing external body movements (kinematics), since the internal, movement-generating forces (kinetics) are not directly accessible [7]. However, the underlying forces are equally important for understanding how different limbs coordinate and interact with the environment. Although kinetic relationships can be estimated from kinematic data using biomechanical models, this process is often more difficult and less reliable in individuals with impairments.

The only possibility to directly measure forces applied to the body, are interaction forces with the environment. During standing and gait of individuals without physical impairments, the foot sole is the only body part that establishes ground contact. Therefore, this plantar interface is of high relevance for determining kinetics and understanding the biomechanics of gait. In biomechanics, plantar forces are generally termed

Received 18 September 2025; revised 24 November 2025; accepted 25 November 2025. Date of publication 5 December 2025; date of current version 2 February 2026. This work was supported in part by the Deutsche Forschungsgemeinschaft (DFG) under Grant 509096131 and Grant 450821862. The associate editor coordinating the review of this article and approving it for publication was Dr. Wei Tang. *(Corresponding author: Bastian Latsch.)*

Bastian Latsch, Felix Herbst, Mark Suppelt, Julian Seiler, Stephan Schaumann, Sven Suppelt, Alexander A. Altmann, and Mario Kupnik are with the Measurement and Sensor Technology Group, Technische Universität Darmstadt, 64283 Darmstadt, Germany (e-mail: mario.kupnik@tu-darmstadt.de).

Martin Grimmer is with the Control and Cyber-Physical Systems Laboratory, Technische Universität Darmstadt, 64283 Darmstadt, Germany.

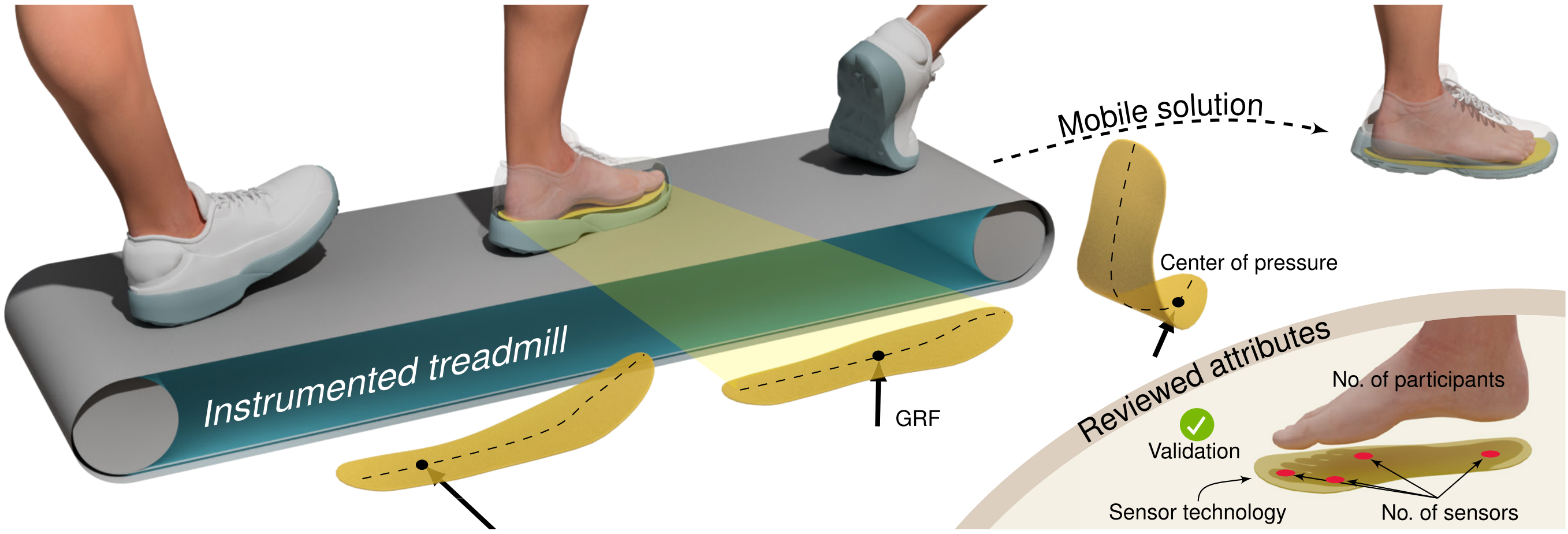

Fig. 1. Ground reaction forces (GRFs) and the center of pressure are fundamental characteristics in gait analysis. Sensor insoles enable the transition from stationary laboratory infrastructure, such as instrumented treadmills, toward mobile, wearable acquisition systems for plantar force and pressure measurements in more natural settings. The individual sensor prototypes differ in several reviewed attributes, such as sensor technologies, number of sensors, number of involved study participants, and validation approaches.

ground reaction forces (GRFs), which can be assessed either on the ground side (floor-based) or on the body side (in-shoe). GRFs are of particular interest for identifying kinematic and kinetic relationships to validate movement models [8]. Humans utilize the GRF-related output of their plantar cutaneous mechanoreceptors as sensory feedback to control movements and maintain postural stability [9], [10]. For instance, balance depends directly on the center of pressure (COP) beneath the foot sole, which is an indicator of the GRF vector direction. Consequently, assessing the pressure distribution during ambulation through pedobarography is one essential part for understanding body movements and indicating abnormalities.

The current gold standard to measure GRFs is the use of force plates, either embedded in the floor or integrated into instrumented treadmills. While floor-embedded force plates cover only individual steps, instrumented treadmills are feasible to observe continuous gait. Although instrumented treadmills and force plates offer higher accuracy than competing body-worn systems, their reliance on stationary laboratory setups can constrain the natural movements and behavior of study participants [11]. In laboratory studies, it is common practice to divert participant attention, which minimizes their awareness and induces more natural behavior [12], [13], [14]. However, the most effective way to observe participants unaffected by experimental conditions is to integrate the clinical study procedures into the daily tasks and capture data unobtrusively in the background. Furthermore, mobile assessment opens up the possibility to analyze realistically complex movements, which contain more than just straight and level walking. Wearable sensor solutions promise to be a major step toward achieving a fully mobile laboratory.

Because the foot sole provides the only ground contact during standing and gait, mobile pressure assessment is commonly conducted using sensor insoles, especially when stationary lab equipment restricts experimental flexibility (Fig. 1). In addition, sensor insoles offer a practical solution for evaluating plantar pressure distribution beneath the foot during ambulation, which ordinary force plates cannot capture, as they only measure total GRFs. Observing pressure distribution is particularly useful in diabetic patients, where small areas under large pressure can lead to ulceration [15], [16]. Other relevant medical conditions motivating mobile gait assessment include but are not limited to Parkinson's disease, stroke-related hemiplegia, and flatfoot condition [17]. However, placing sensors beneath the foot brings its own set of challenges, as they must withstand full body weight and the humid environment of the foot, while integrating seamlessly with the sole to avoid influencing natural gait.

Flexible materials have gained special attention in human-centered applications due to their ability to conform to the body's contours and surfaces. Various sensor technologies and topologies have been extensively studied, with flexible film sensors, in particular, benefiting from ongoing advancements in material science.

Sensor insoles are the focus of current research across multiple disciplines, such as material science, medical engineering, and biomechanics. A wide variety of approaches exists, often shaped by the target application and the research community behind their development. Some insole systems demonstrate high measurement accuracy but lack user-centered considerations, while others introduce novel materials yet fall short in characterization and subsequent human studies. This diversity has led to a fragmented landscape with limited cross-comparability.

### A. Technology Overview

The historical development of sensor insoles, eventually emerging as commercially available systems, provides the context for identifying current shortcomings and the contribution of this review. Original research in scholarly works is then examined in depth, organized by their underlying transducer principles, enabling a comparative analysis of their features, capabilities, and experimental procedures used.

Mobile foot pressure assessment has been available for centuries and continues to evolve. One of the

TABLE I

SELECTION OF COMMERCIALLY AVAILABLE SENSOR INSOLES FOR PLANTAR PRESSURE MEASUREMENTS, USING RESISTIVE (R) OR CAPACITIVE (C) SENSOR TECHNOLOGIES AND WLAN OR BLE FOR CONNECTIVITY. PRODUCTS ARE ORDERED ALPHABETICALLY

| Product | Manufacturer | No. of sensors | Communication | Sampling (Hz) | R | C |
|---|---|---|---|---|---|---|
| F-Scan GO [18] | Tekscan (Boston, MA, USA) | 966 | WLAN | 500 | × | |
| Intelligent Insoles Pro [19] | XSENSOR (Calgary, AB, Canada) | 235 | BLE | 150 | | × |
| Loadsol [20] | Novel (Munich, Germany) | 3 | BLE | 200 | | × |
| Medilogic insole [21] | Medilogic (Schönefeld, Germany) | 240 | WLAN | 400 | × | |
| Pedar [22] | Novel (Munich, Germany) | 99 | WLAN | 400 | | × |
| ReGo [23] | Moticon (Munich, Germany) | 16 | BLE | 100 | | × |

earliest sensor-integrated insole prototypes was presented by Pedotti et al. [24]. It consists of a piezoelectric polyvinylidene fluoride (PVDF) sheet with 16 aluminum electrodes. Among the first commercial insoles were the resistive F-Scan (Tekscan, Boston, MA, USA) and the capacitive Pedar (Novel, Munich, Germany) systems, introduced in the early 1990s and still available today in revised forms. A review by Cavanagh et al. [25] from that time provides an overview of several existing approaches in scholarly works. In order to evaluate their performance, most custom-built sensor insoles in research are compared to commercial force plates, which are considered the gold standard for GRF assessment. When force plates are not feasible, such as in the case of mobile experimental setups, commercial insoles are often taken as a reference for the validation.

Typically, commercial insole systems utilize wireless local area network (WLAN) connectivity when higher sampling rates and a large number of sensors are specified. Power-efficient Bluetooth Low Energy (BLE) is commonly employed when data rate is of less concern. In insole systems, including commercial insoles, users must generally accept lower accuracy compared to stationary measurement equipment due to space and weight limitations inherent in wearable sensor systems. However, when this tradeoff is acceptable or required for the measurement setup, sensor insoles allow for experiments to be conducted outside of the laboratory environment.

Most commercially available insoles rely on resistive or capacitive sensor technologies [26]. The most current versions of widely used insoles include those from Tekscan, Novel, Moticon, XSENSOR, and Medilogic (Table I). Novel's Pedar system and Tekscan's F-Scan insole have evolved into numerous variants since their first introduction in the 1990s. They are widely used in clinical studies [27], [28], [29], [30] and are often referred to as reference systems for custom-built insoles [31], [32], [33], [34], [35]. Novel's more recent Loadsol system features accurate GRF measurement capabilities rather than high-resolution pressure distribution [20]. It offers up to three segmented regions for force measurement and includes small-form-factor electronics fixed to the shoe for portability. Most commercial insoles and custom-built prototypes attach the electronics near the ankle or distal shank to minimize interference with foot movement. Notably, Moticon's commercial systems, OpenGo and ReGo [23], feature electronics fully embedded into the insole, making these systems more user-friendly and largely unobtrusive to the user (Fig. 2).

Medilogic insoles feature a measurement range up to $64\,\mathrm{N\,cm^{-2}}$ with 240 resistive sensors at 400-Hz sampling rate [21]. One of the highest spatial resolutions available with BLE technology is offered by XSENSOR (Calgary, AB, Canada), providing up to 235 sensor elements across the entire footbed [19]. Their Intelligent Insoles Pro system measures pressures up to $88\,\mathrm{N\,cm^{-2}}$ with a resolution up to 7 mm, depending on the shoe size. Several commercial insole systems that were previously reported in other publications [36], such as BioFoot, seem to have left the market and are no longer available. BioFoot was one of the few commercial piezoelectric insole systems.

While many commercial insole systems focus primarily on measuring normal pressures, it is increasingly recognized that shear stress distribution beneath the foot sole may play an important role in the development of diabetic foot ulceration [15], [37]. However, little is known about the influence of normal and shear stress distribution, as no system currently exists that can assess such data on a broad scale. Force plates can acquire three-dimensional force components but lack the spatial resolution required to detect pressure hotspots. Hotspots can lead to skin and tissue breakdown in the diabetic foot. Sensor-integrated shoes and insoles offer a feasible solution for assessing the pressure distribution using multiple sensor elements. Until a few years ago, three-dimensional force acquisition was either unavailable in mobile solutions or required bulky transducers beneath the shoe [38], [39]. Veltink et al. [38] use two 6-degrees-of-freedom force and moment sensors under the shoe to calculate the COP [Fig. 3(a)]. Sorrentino et al. [40] employ such load cells as a reference to verify their capacitive insole [Fig. 3(b)]. Both prototypes have in common that the shoe sole is thicker and heavier than normal due to the introduction of load cells and appears rather bulky.

However, recent advances in the production and handling of polymer-based materials and sensors gave rise to several thin insole prototypes in research that incorporate shear-force measurements [41], [42]. At the time of composing this review, no commercial insole systems capable of spatially resolved shear force measurements are available to the best of our knowledge. A recent literature review came to the same conclusion [17]. Most insoles only estimate the shear forces from multiple sensor elements, which prevents the assessment of high-resolution maps. However, such high resolution is particularly important for diabetic patients who are at risk of foot ulceration, as identifying areas of elevated plantar pressure may help predict ulcer development [16]. Consequently, this topic holds considerable potential for future research.

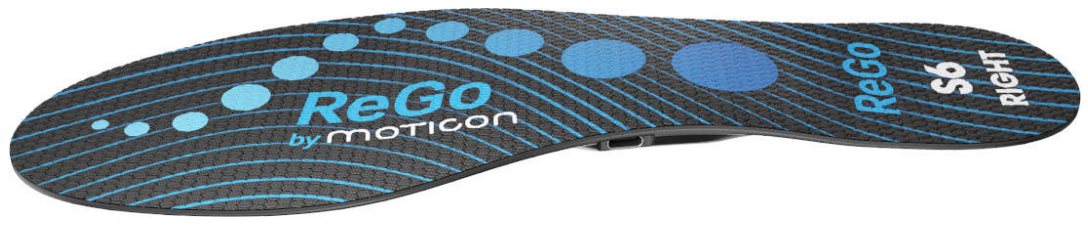

Fig. 2. State-of-the-art commercial sensor insole (ReGo, Moticon, Munich, Germany) with fully embedded electronics in the arch of the foot area, designed for maximum unobtrusiveness for the user. Reproduced, with permission from Moticon. Copyright Moticon.

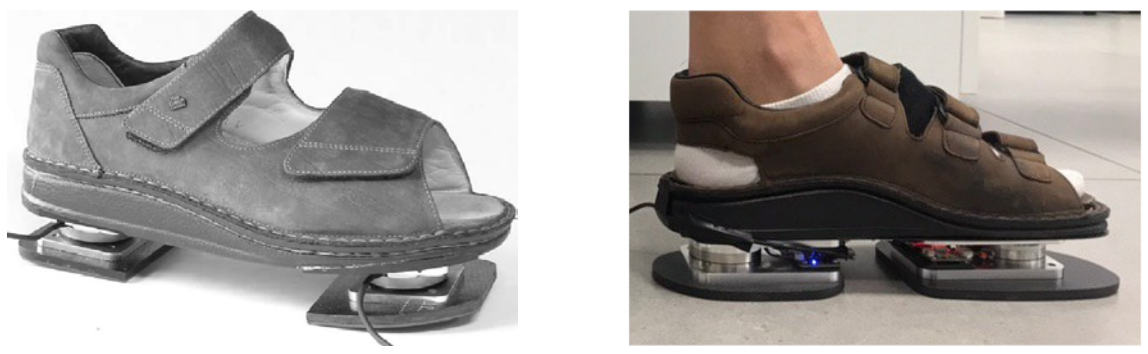

(a) (b)

Fig. 3. Load cells integrated beneath shoes enable multiaxial force measurement but account for a rather bulky appearance. (a) Two 6-degrees-of-freedom force and moment sensors positioned beneath the shoe facilitate multiaxial GRF measurements. Reproduced, with permission, from [38]. Copyright 2005, IEEE. (b) Similar instrumented shoe utilized as a mobile reference system for calibrating a sensor insole. Picture cropped in accordance with the CC BY 4.0 license, from [40].

Several other medical conditions also benefit from the development of sensor insoles for mobile gait assessment. Chen et al. [17] provide an overview of relevant examples, including diabetic foot ulcers, Parkinson's disease, stroke-related hemiplegia, fall risk in older adults, and flatfoot condition. Sensor insoles show promise across these conditions, and ongoing advances in sensing materials, electronics, and gait analysis are expected to yield increasingly capable and clinically relevant systems in the near future.

### B. Overview and Limitations of Prior Reviews

Commercial insoles typically do not reflect the latest state of the art, as their development requires years of continuous work. In contrast, prototypes presented in scholarly works feature rather unexplored technologies that open up completely new opportunities. In addition, academic efforts to develop custom systems are often driven by the limitations of commercial systems, including high costs. Several reviews of sensor insoles are available in the literature, each focusing on different aspects. Ramirez-Bautista et al. [36] focus on disorders and injury detection, while giving an overview of commercially available insoles along with several research prototypes. Wang et al. [43] report on multiaxial sensors for the prevention of diabetic foot ulceration. Prasanth et al. [44] present wearable sensors capable of real-time gait analysis, while Subramaniam et al. [45] provide an overview of insole systems that also extend to heart-rate and general health monitoring. Eskofier et al. [46] focus their overview on smart shoes in the context of the Internet of Things, emphasizing their role in pathological diagnostics and treatment monitoring, while also deriving generally applicable conclusions regarding technological requirements and engineering gaps in medical foot-related applications.

Chen et al. [17] provide a comprehensive review about sensor technologies, read-out electronics, signal processing, meaningful data extraction, and their applications in disease analysis. The authors emphasize that the accuracy of subsequent signal processing and disease identification ultimately depends on the quality of the raw sensor data. They identify sensor slip relative to the foot as a critical source of low data quality. However, they provide no concrete recommendations for sensor validation. In our view, these represent two distinct problems that each require targeted solutions rather than one being prioritized over the other. Moreover, the issue of slippage can be extended and generalized to the broader challenge of achieving deterministic sensor behavior in contact with the human body, which requires a more detailed analysis, as discussed in Section IV-E.

Errors in classification rendered some reviews partially unhelpful and potentially misleading. A broad overview from commercial insoles to research prototypes and their applications is given by Almuteb et al. [47]. However, the section on plantar pressure sensors lacks focus on the underlying sensor technology and contains misclassifications. For instance, most sensor insoles categorized as piezoelectric in Tab. 2 of their review [47] are, in fact, not piezoelectric but utilized other sensor technologies. It remains unclear whether ambiguous descriptions in the primary works led to the incorrect classification or if other factors were involved.

In general, most reviews containing sensor insoles focus on application-oriented selection and categorization of scholarly works, with less emphasis on the sensor technology itself. While they offer insights into measurement technologies broadly applicable to gait analysis, they lack a deeper examination of the specific sensor technologies used in insoles and their fundamental validation to increase the quality of recorded data.

## II. Scope and Review Methodology

By critically evaluating recent advances in sensor technologies, this review aims to support future development of innovative, user-oriented sensor insoles and their appropriate characterization for accurate pressure analysis beneath the foot. We will focus on sensor characteristics and evaluation methods with respect to multimodal sensing, force and pressure measurement capabilities, sensor number and placement, multiaxial measurements, and validation approaches (Fig. 1).

The sensor insoles presented in Section III are organized by their primary underlying transducer principle. As with commercial insoles, resistive and capacitive sensors are the most prevalent. However, inductive, piezoelectric, and less common technologies are also included in this literature review, as these sensors are relatively unexplored in insoles and may be worth considering for future investigations. Inertial measurement systems are intentionally excluded from this review, as they provide only indirect pressure estimates derived from kinematic relationships [48], [49]. Furthermore, with the exception of rare implementations at the foot sole [50], inertial measurement units (IMUs) are predominantly positioned on the foot or leg [51], [52], [53].

The selection of references in this review was guided by relevance and informational depth rather than strict adherence to systematic review protocols due to the heterogeneity of the existing literature. In particular, a substantial portion of studies utilizes resistive FSR sensors in a conventional manner, offering little to no innovation in sensor or study design. Conversely, other publications present sensors highly engineered from a materials science perspective, with the insole application only as a peripheral motivation. This divergence in focus between generic use and limited application relevance makes it difficult to define objective inclusion criteria. As a result, a subjective prescreening step was required during the review of several hundred original works to ensure a focused and meaningful synthesis of the topic. We started with works known to us, frequently cited works, and works mentioned in the review papers referenced in Section I-B. Subsequently, references cited within these works and papers citing them were examined, including more recent works from the same research groups.

In total, 184 related publications were cataloged and screened in detail, focusing on original designs in works of acceptable quality that preferably incorporated the insole's application in human experiments or featured otherwise unique characteristics. Table II provides an overview of 41 selected state-of-the-art sensor insoles for plantar pressure measurements, deemed most relevant by the authors based on their fundamental differences and their novelty at the time, with the aim of keeping the list as concise as possible. This list is evaluated in Section III and exclusively includes custom insole prototypes designed for humans with two biological legs, which are developed in scholarly works; commercial insoles are not included. The selection contains 16 resistive, 7 capacitive, 2 inductive, 9 piezoelectric, 2 triboelectric, 2 optical, and 3 multimodal sensor insoles. However, it is not feasible to draw conclusions about the proportional representation of transducer principles in the literature from this sample, given the selection process used. Nevertheless, due to the combined quality and relevance of the included works, this review can be regarded as a key reference in the field.

## III. Sensor Transduction Principles

The sensor technologies are listed in order of passive property changes (resistive, capacitive, and inductive), active transducers (piezoelectric and triboelectric), and optical methods, with each group arranged by decreasing prevalence. Based on this comprehensive literature review, requirements for insole prototypes are derived, and general conclusions on their applications are drawn (Section V).

### A. Resistive

Most custom-built insoles utilize commercially available resistive sensors (Fig. 4). These resistive insoles are often based on force-sensing resistors (FSRs) [64], [84], [89], [90], [91]. Commercial FSRs consist of conductive layers in which applied pressure increases the real contact area between conductive particles and electrodes, thereby reducing the overall sensor resistance. Early FSR versions could only detect on and off states, essentially functioning as foot switches [25]. However, progress in manufacturing technologies gave rise to commercial sensors with more defined characteristics, enabling continuous, analog sensor outputs.

Among the earliest uses of FSRs in clinical studies is a study by Smith et al. [90], which evaluates them as triggers for functional electrical stimulation in cerebral palsy patients. However, 5.5% of the performed steps are not detected at all, with 80% of these missed due to a smaller-than-expected signal below the detection threshold programmed. Bamberg et al. [88] are among the first to develop a wireless wearable system that employs numerous sensor modalities, including FSR sensors along with PVDF-based sensors and an IMU. However, the piezoelectric PVDF sensors seem not to be incorporated into the gait analysis concept, leaving inertial and resistive data as the main inputs for the sensor fusion. González et al. [84] use four FSR sensors placed beneath an insole for gait monitoring [Fig. 4(a)]. Choi et al. [33] compare COP trajectories from a six-FSR insole with those from the F-Scan system by having participants wear both simultaneously. Lim et al. [32] employ a similar setup to detect gait phases for synchronous exoskeleton control. Park et al. [68] use five FSRs to additionally detect the timing of weight transfer between feet. Khandakar et al. [62] adhere 16 FSRs onto an insole to analyze the gait cycle of twelve participants. The presented variability of one participant's 10-meter walk is relatively large. As no reference system is utilized, it remains unclear if the walking itself was variable or the sensor behavior. In a previous work [35], some of the authors used the commercial F-Scan system as a reference, however, not concurrently to the FSR insole but subsequently. Negi et al. [64] utilize two FSR sensors below the heel and toes to detect gait phases, while additional electromyography electrodes at the tibialis anterior and gastrocnemius assess the muscle activity, and a shank-attached IMU measures kinematics. Roth et al. [78] embed three FSR sensors, the read-out electronics, the battery, and a coil for wireless energy transfer into an insole to build a fully integrated system for home monitoring.

Other methods include polymeric films impregnated with carbon black, known as Velostat or Linqstat, and other piezoresistive materials instead of resistive FSR sensors [70], [92], [93]. Saito et al. [31] employ pressure-sensitive rubbers to measure the plantar pressure beneath the feet of one young adult and two elderly people. Huang et al. [94] utilize carbon black-doped, porous thermoplastic polyurethane (TPU) as a piezoresistive layer on a flexible printed circuit board (PCB). The active layer is structured into 32 sensor elements, which have been presented in previous works [95], [96]. In addition, the insole contains an ethylene-vinyl acetate (EVA) layer, which is employed as a triboelectric nanogenerator to harvest energy for the electronics. The insole is tested with one participant on a subjective basis. Tan et al. [86] employ a row-column electrode configuration to assess 75 sensor elements using only 20 electrode strips, with a piezoresistive rubber in between.

TABLE II

State-of-the-Art Noncommercial Sensor Insoles for Plantar Pressure Measurements, Using Numerous Sensor Technologies: Resistive (R), Capacitive (C), Inductive (L), Piezoelectric (Q), Triboelectric (T), and Optical (O).The Study Setups Are Specified With the Physical Reference Used for Sensor Characterization and the Reference System Used for Gait Experiments. The Number of Patients Is a Subset of the Total Participants

| Scholarly work | Year | No. of sensors* | Sensor reference | No. of participants | No. of patients | Gait reference | R | C | L | Q | T | O |
|---|---|---|---|---|---|---|---|---|---|---|---|---|
| Willemstein et al. [54] | 2025 | 4 | Testing machine | 9 | 0 | Instrumented treadmill | × | | | | | |
| Dai et al. [55] | 2024 | 13 | None | 4 | 0 | Force plate | | × | | | | |
| Hu et al. [56] | 2024 | 4 | Testing machine with inclination | 1 | 0 | None | | | | | × | |
| Latsch et al. [57] | 2024 | 4 | Testing machine | 1 | 0 | Instrumented treadmill | | | | × | | |
| Luna-Perejón et al. [58] | 2023 | 12 | Multiple weights | 1 | 0 | None/Custom FSR insole | | × | | | | |
| Tang et al. [59] | 2023 | 4 | Testing machine | 6 | 5 | XSENSOR insole | | × | | | | |
| Yabu et al. [60] | 2023 | 4/4 | Testing machine | 11 | 0 | Instrumented treadmill | × | | | × | | |
| Ben Dali et al. [61] | 2022 | 8 | One weight | 1 | 0 | None | | | | × | | |
| Khandakar et al. [62] | 2022 | 16 | Multiple weights | 12 | 0 | None | × | | | | | |
| Wang et al. [42] | 2022 | 64 | Shear testing machine | 3 | 0 | None | | | × | | | |
| Chen et al. [63] | 2021 | 24 | Load cell | 2 | 1 | None | | | | × | | |
| Gao et al. [41] | 2021 | 32 | Unknown | 1 | 0 | None | | | | × | | |
| Negi et al. [64] | 2021 | 2 | Multiple weights | 6 | 0 | None | × | | | | | |
| Prado et al. [65] | 2021 | 3 | None/Raw data usage | 10 | 10 | Sensorized walkway | × | | | | | |
| Zhao et al. [66] | 2021 | 48 | None/Raw data usage | 10 | 0 | None | × | | | | | |
| Hao et al. [67] | 2020 | 4 | None | 1 | 0 | None | | | | | | × |
| Park et al. [68] | 2020 | 5 | None | 8 | 0 | None/Literature | × | | | | | |
| Sorrentino et al. [40] | 2020 | 280 | Pneumatic pressure | 1 | 0 | Load cell shoes | | × | | | | |
| Tao et al. [69] | 2020 | 24 | Testing machine | 1 | 0 | None | | × | | | | |
| Acharya et al. [70] | 2019 | 8 | Testing machine | 0 | 0 | No participants | × | | | | | |
| Chandel et al. [71] | 2019 | 5 | None | 5 | 0 | None | | | | × | | |
| Jasiewicz et al. [72] | 2019 | 8 | Electrodynamic testing machine | 20 | 0 | None/Static only | | | | × | | |
| Lin et al. [73] | 2019 | 2 | Load cell | 1 | 0 | None | | | | | × | |
| Nie et al. [74] | 2019 | 8 | Load cell | 1 | 0 | None | | | × | | | |
| Zhu et al. [75] | 2019 | 4/5 | Testing machine | 3 | 0 | None | | | | × | × | |
| Chen et al. [76] | 2018 | 48 | One weight | 2 | 0 | None | × | | | | | |
| Choi et al. [33] | 2018 | 6 | None | 8 | 0 | F-Scan insole | × | | | | | |
| Deng et al. [77] | 2018 | 32 | Testing machine | 0 | 0 | No participants | | | | × | | |
| Roth et al. [78] | 2018 | 3 | None | 15 | 15 | Sensorized walkway | × | | | | | |
| Lim et al. [32] | 2017 | 4 | None | 5 | 0 | F-Scan insole | × | | | | | |
| Rajala et al. [79] | 2017 | 8 | Load cell | 5 | 0 | None | | | | × | | |
| Vilarinho et al. [80] | 2017 | 5 | Testing machine | 1 | 0 | None | | | | | | × |
| Han et al. [81] | 2016 | 2 | None | 3 | 0 | None | | | | × | | |
| Lin et al. [82] | 2016 | 48 | None | 2 | 0 | None/Step counter | × | | | | | |
| Ashad Mustufa et al. [83] | 2015 | 32 | None | 0 | 0 | No participants | | × | | | | |
| González et al. [84] | 2015 | 4 | One weight | 5 | 0 | None/Literature | × | | | | | |
| Motha et al. [85] | 2015 | 3 | Load cell | 1 | 0 | None | | × | | | | |
| Tan et al. [86] | 2015 | 75 | Force plate, unknown loading | 1 | 0 | Force plate | × | | | | | |
| Saito et al. [31] | 2011 | 7 | Load cell | 3 | 2 | F-Scan insole | × | | | | | |
| Shu et al. [87] | 2010 | 6 | Unknown | 8 | 0 | Force plate | × | | | | | |
| Bamberg et al. [88] | 2008 | 4/2 | Testing machine | 15 | 5 | Force plate | × | | | × | | |

*For entries A/B with multiple sensor technologies, A is associated with the leftmost marked technology column in that row, and B with the rightmost marked column.

One sensor is calibrated using a force plate as a reference, however, the method of force application is not specified. The results show substantial hysteresis and drift, yet the system is still able to estimate a 700-newton body weight during still standing of one participant, with a reported standard deviation of 70 N.

Anzai et al. [34] employ pressure-sensitive conductive rubber in seven sensor elements, distributed across the insole. With 13 healthy participants, the COP trajectories during normal walking are compared to the simultaneously acquired data from the commercial F-Scan insole that is placed on top of the prototype. The Pearson's correlation coefficients of up to 0.99 indicate a linear relationship between both insoles, with greater correlation on the anterior-posterior axis than the medial-lateral axis.

Zhao et al. [66] utilize 48 sensors based on piezoresistive ink for assessing plantar stress distribution and recognizing locomotion modes [Fig. 4(d)]. Willemstein et al. [54] use conductive pellets in a 3D printer to fabricate foam-like piezoresistive sensors through a process referred to as liquid rope coiling [Fig. 4(e)]. The sensors are characterized for hysteresis over 10 000 load cycles with test forces up to 800 N. Using four sensors placed across the plantar surface, they apply personalized Hammerstein–Wiener models to estimate all three GRF components during normal-speed walking of nine participants on an instrumented treadmill. Although the accuracy of the estimates likely remains limited to walking and may not be generalized to other forms of locomotion, the study demonstrates that analytical estimation of GRF components is feasible with a small number of sensors. Similar

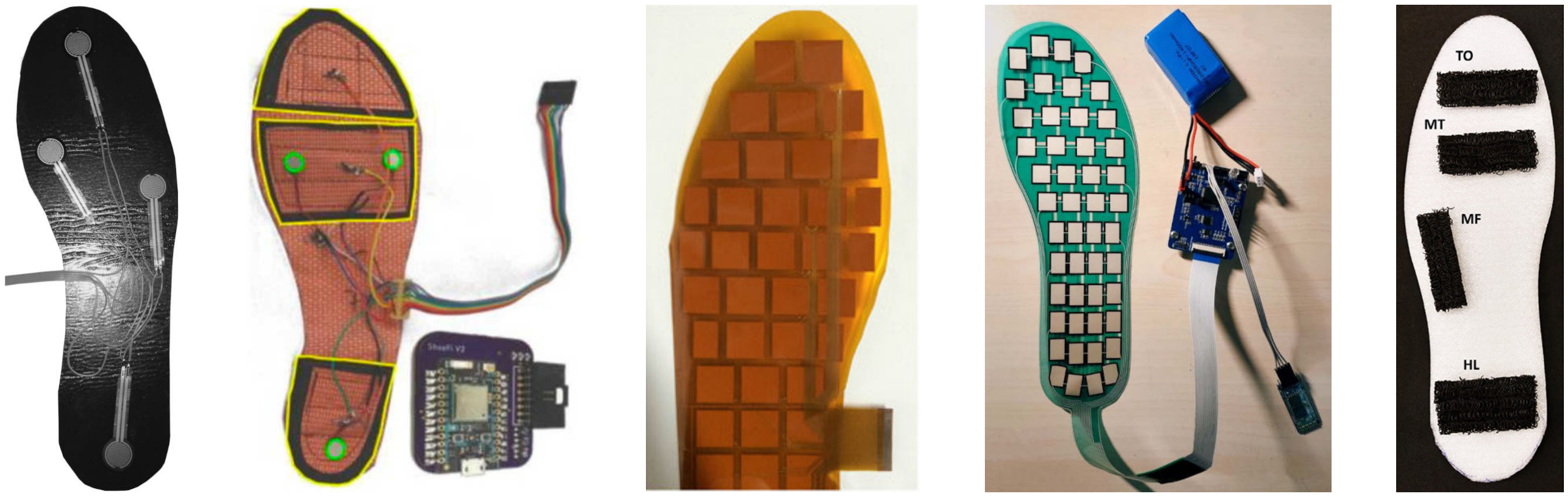

(a) (b) (c) (d) (e)

Fig. 4. Resistive sensor insoles are often based on FSR sensors or conductive flexible films. (a) González et al. [84] build a resistive sensor insole with four FSR sensor elements. Reproduced in accordance with the CC BY 4.0 license, from [84]. (b) Prado et al. [65] utilize piezoresistive fabric between two copper fabric layers to detect freezing of gait in ten patients with Parkinson's disease. In addition, the insole comprises three vibration motors (green) to possibly release the patient from the freezing state. Reproduced with permission, from [65]. Copyright 2021, IEEE. (c) Lin et al. [82] utilize a piezoresistive textile composed of yarn coated with a piezoelectric polymer to create a matrix insole, incorporating 48 sensors that cover 80% of the foot's sole area. Reproduced with permission, from [82]. Copyright 2016, IEEE. (d) Zhao et al. [66] utilize 48 sensors based on piezoresistive ink for assessing plantar stress distribution and recognizing locomotion modes. Reproduced with permission, from [66]. Copyright 2021, IEEE. (e) Willemstein et al. [54] 3D print four piezoresistive foam sensors to estimate all three GRF components during walking on an instrumented treadmill. Reproduced in accordance with the CC BY 4.0 license, from [54].

findings reported in the literature mostly use machine learning approaches [97], [98].

Prado et al. [65] utilize piezoresistive fabric between two copper fabric layers to detect freezing of gait in ten patients with Parkinson's disease [Fig. 4(b)]. In addition, the insole comprises three vibration motors to possibly release the patient from the freezing state. Using neural networks, the authors detect gait phases [99] and spatiotemporal gait characteristics, such as stride length, stride width, and stride time [65]. The model uses raw data without any filtering from the piezoresistive sensor insole and one IMU placed on the dorsum of the foot to estimate these parameters. Without preprocessing, a rapid mapping of data is possible and has the potential to enable real-time identification in the future. Shu et al. [87] estimate the participant's body weight using a piezoresistive strain-sensing fabric, achieving an average deviation of 6.9%. The sensor's accuracy and zero drift are each specified at 5%. The measurement range is reported as 10 to 1000 kPa with a resolution of 1 kPa. However, the test setup used for sensor characterization and the origin of the reported system properties are not described.

Lin et al. [82] utilize a piezoresistive textile composed of yarn coated with a piezoelectric polymer to create a matrix insole, incorporating 48 sensors that cover 80% of the foot's sole area [Fig. 4(c)]. They compute the numerical derivative of the sensor voltage to independently detect heel strike and toe-off events, irrespective of voltage offsets or participant weights. Chen et al. [76] build upon the work of Lin et al. [82] and embed commercially available piezoresistive fabric between two flexible PCBs for contacting up to 48 sensor elements. Their insole is designed as a single template and later cut to customize the outline for individual foot sizes. The common ground electrode on the bottom of the insole is intended to reduce crosstalk, which the authors report with row-column electrode designs, such as those described by Tan et al. [86]. The piezoresistive sensors are not individually calibrated. Instead, the bulk material is loaded with 130 kPa at different locations. The resistance is reported to be around 50 Ω, but it remains unclear whether this value is differential or absolute, and how the sensor performs in terms of linearity and hysteresis. A gait analysis experiment with two participants qualitatively identifies pressure hotspots. However, these are not verified against reference systems. In a subsequent study [100], the authors compare the sensor insole to a force plate during walking and running. After normalizing the insole and force plate values to their maxima for each step, they report a correlation of 0.989. This normalization, however, cancels out interstep variations in sensor amplitude, meaning the correlation reflects only the shape of each step signal. Nevertheless, the shape appears sufficient for motion classification, achieving accuracies above 99.6%.

## B. Capacitive

Capacitive sensors for plantar pressure measurement use a compressible dielectric material between two conducting plates. Most commercially available insoles, such as Novel's Pedar [30], Novel's Loadsol [101], and Moticon's OpenGo [102] system, rely on the capacitive operating principle. Capacitive insole measurements feature high accuracy and high resolution, whereas their susceptibility to electromagnetic interference (EMI) is a major downside [17]. Furthermore, intruding humidity and temperature dependence are of particular concern in capacitive sensors placed beneath the foot. It appears that a few elastic dielectric materials, which are less affected by these factors, and compensation methods have been included into patents by market-leading companies [103], [104].

Other solutions include interdigitated sensors embedded into polydimethylsiloxane (PDMS) [85] and layered electrodes with elastic dielectrics in between, such as PDMS [58] and other silicone rubbers [69]. Tao et al. [69] demonstrate increased sensor sensitivity by laser-cutting pores into a silicone rubber dielectric. 3D printing of capacitive sensors for the application in insoles is also a feasible approach for fast customizability [105]. Tamm et al. [106] present an insole prototype based on flexible PCBs, which was relatively advanced at that time. They describe the PCB to be suitable for capacitive as well as resistive measurements. Pressure is applied on a dynamographic platform with PDMS foam deployed as a dielectric for capacitive assessment. Ashad Mustufa et al. [83] present an insole with 32 sensor elements on a flexible PCB. However, no system validation nor testing with participants is conducted.

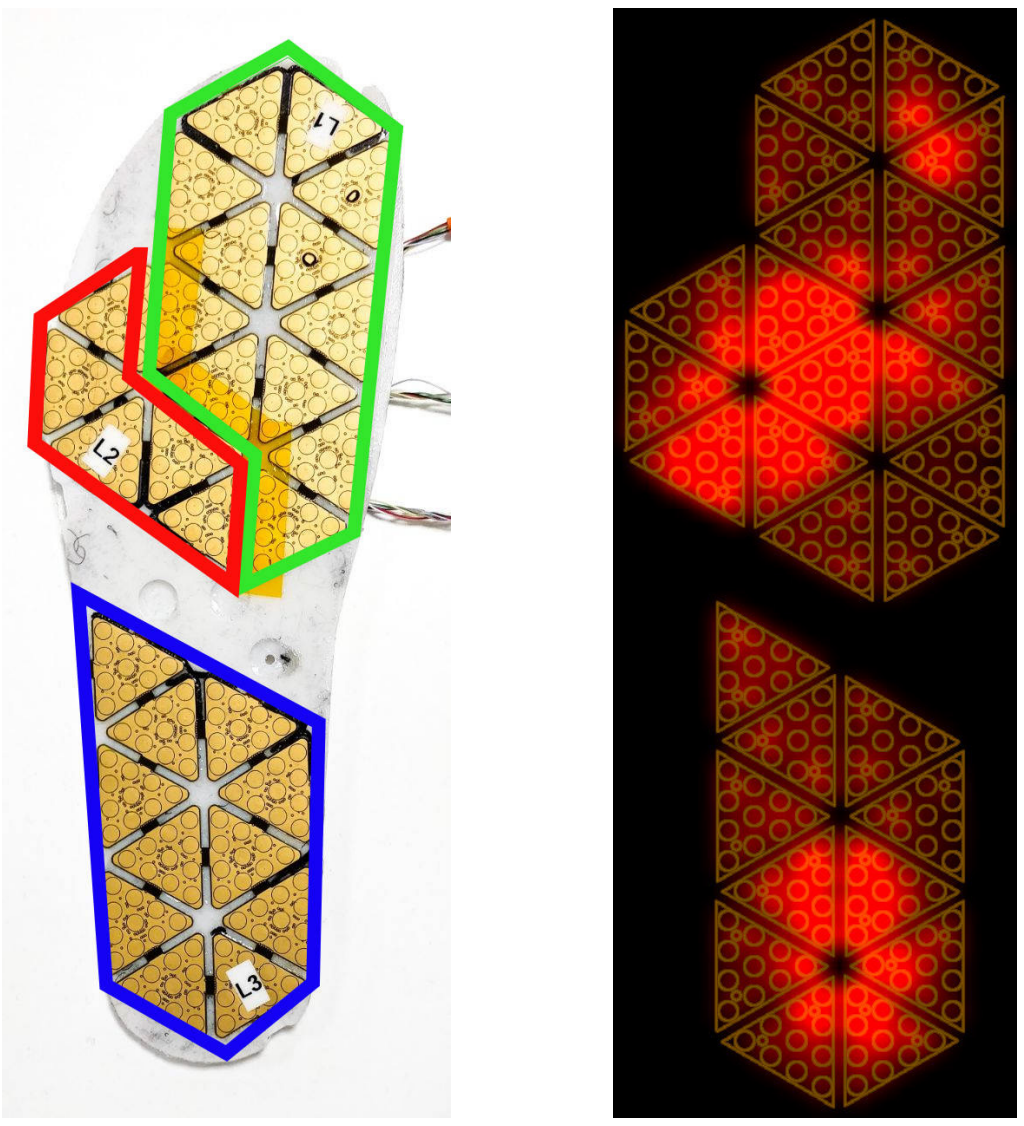

(a) (b)

Fig. 5. Capacitive sensor insole using large-scale sensor arrays. (a) Insole is split into three patches with a total of 28 segments to measure the pressure distribution beneath the foot sole with high resolution. (b) Bright, red areas on the heat map indicate the highest pressure regions. Both pictures reproduced in accordance with the CC BY 4.0 license, from [40].

Tang et al. [59] employ their four-sensor insole on diabetic patients and compare peak pressure to those measured by the commercial XSENSOR system. In addition, the triaxial sensor from a previous work [107] is characterized using a testing machine, assessing its response to normal pressures up to 300 kPa and shear forces up to 90 kPa under both static and 1-Hz dynamic loading profiles.

Sorrentino et al. [40] utilize large-scale capacitive sensor arrays, originally developed for tactile sensing in humanoid robots [108], which consist of ten elements per sensor segment. Three patches with a total of 28 segments are integrated into an insole to measure the pressure distribution beneath the foot sole (Fig. 5). The pressure distribution and COP are compared to a sensorized shoe equipped with conventional load cells [Fig. 3(b)]. Dynamic foot movements during walking produce higher accuracy than during standing still. The authors report a slight capacitance offset without weight on the foot in comparison to the baseline value, which represents the capacitance when the insole is outside the shoe. They attribute this offset to the pressure applied by the foot onto the insole when the shoe is affixed to the foot sole.

Dai et al. [55] explore an approach to overcome the deficits in dynamic responses during jumping by compensating the 13 insole sensor signals with data from an IMU. The forces applied during landing partially exceed 1500 N with all participant masses below 100 kg. An experiment with four participants on a force plate as a reference yields enhanced dynamic output after processing it with the neural network presented. This work demonstrates a possible approach how spatially resolved pressure sensors beneath the foot can be combined with inertial measurement data. Luna-Perejón et al. [58] utilize a separate shield around the sensor electrodes to reduce EMI. The shield is actively driven by the capacitance-to-digital converter, implementing a guarding technique that eliminates parasitic capacitance to ground. Their insole is calibrated using a set of weights and compared to a self-developed FSR insole.

### C. Inductive

Recently, inductive sensor insoles have seen increased use. One of the earliest designs intended for insole integration includes dedicated wire-wound coils, elastic rubber spacers, and a steel plate as the inductive target, resulting in a relatively bulky thickness of over 10 mm [109]. Later prototypes utilize coils on PCBs to reduce size and increase integrability [110]. The recent success of flexible PCBs has drastically reduced costs, facilitating broader availability and promoting their use in prototyping. Most current concepts now employ flexible PCBs to integrate planar coils and measure inductance using $LC$ resonant circuits.

Nie et al. [74] employ ferrite films and planar coils, separated by a textile spacer, to detect sensor deformation through a resonance frequency shift. The coils are designed as part of a near-field communication circuit, which allows for the remote reception of the frequency shift at a distance of a few millimeters. While the remote radio-identification approach is promising, the current operational distance appears insufficient for practical insole readouts in real-world scenarios, where participants are in motion.

Multiple inductance values from adjacent coils can be translated into spatial components using simple geometric calculations [110], [111]. Achieving three-dimensional force acquisition, including shear forces in the plantar plane, requires at least three coils per sensor element. Wang et al. [112] examine various coil shapes and topologies to determine the optimal configuration. They conclude that a design with four square coils outperforms other configurations tested in terms of sensitivity and power efficiency. Based on their previous work, Wang et al. [42] develop an inductive sensor insole. It features 64 triaxial sensor elements capable of measuring two shear force components, in addition to the normal force. They motivate their work by emphasizing the early detection of excessive shear forces on the plantar surface, which may play a critical role in preventing foot ulceration in patients with diabetes [113]. The triaxial sensor elements consist of three stacked coil layers, an elastomer spacer disk, and a metallic target on top (Fig. 6). The metallic targets are positioned within the effective range of the coils, with the elastomer spacers between them, which deform under applied load. As the metallic target moves relative to the coils, it causes individual inductance changes, allowing the sensor to separately measure normal and shear forces through simple geometric calculations. Even though no overall thickness is specified, adding up the individual layers yields an approximate insole thickness of 2 mm, which is thinner than most other triaxial sensor insoles. The elastomer layer accounts for the majority of the thickness at 1.5 mm, and the system is reported as comfortable by participants. Normal plantar pressure is measured at a root mean square error of 2.1%. However, the extensive sensor calibration required due to the large number of sensors is noted as a drawback by the authors. Variations in the insole interface, such as different socks and shoes, lead to discrepancies between participants.

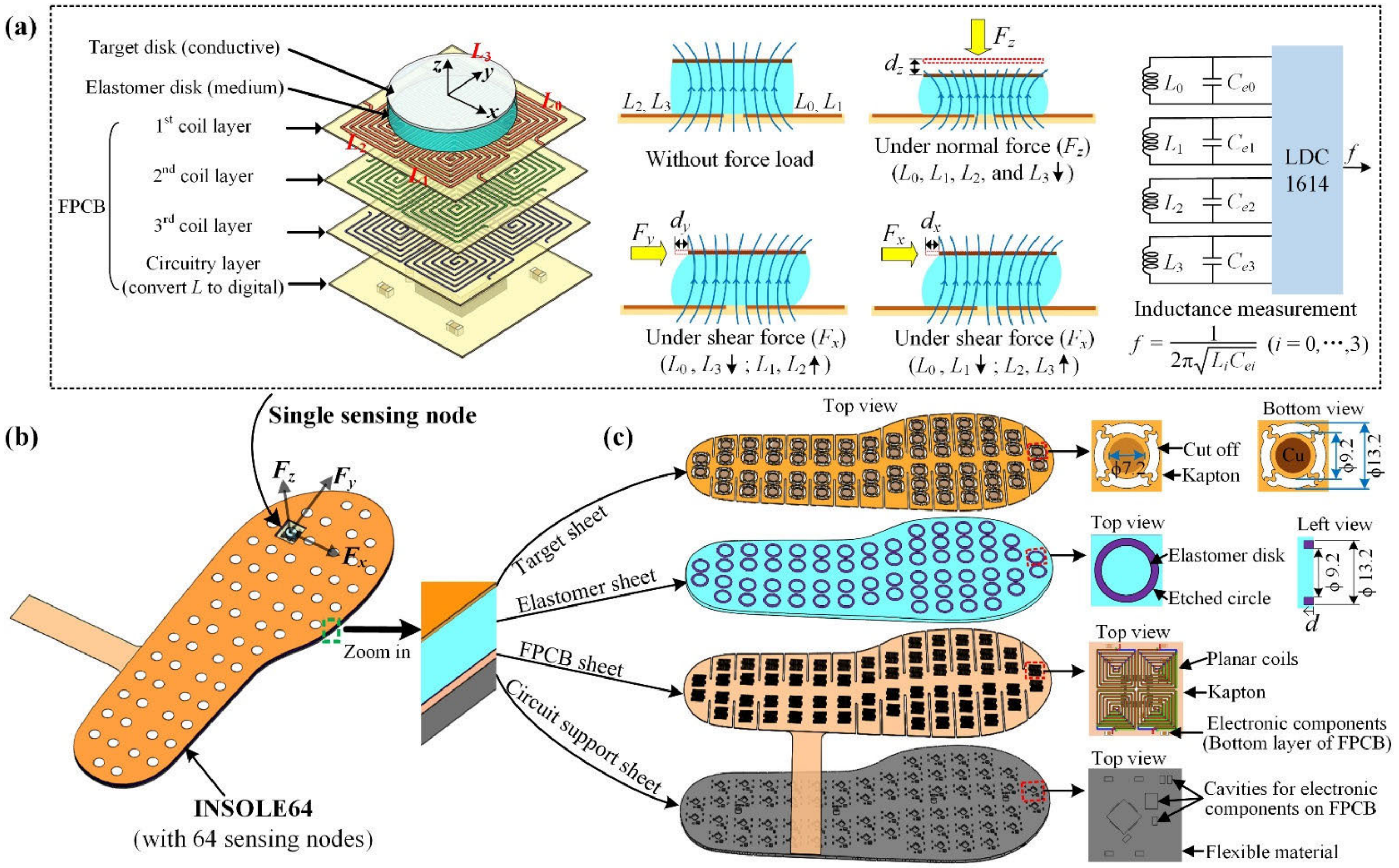

Fig. 6. Inductive sensor insoles are mostly based on measuring the inductance of several coils concurrently, enabling triaxial measuring systems for normal and shear forces. (a) Metallic targets are placed into the effective range of the coils with elastomer spacers in between that are deformed through the force to measure. (b) This setup enables the measurement of all three force components, including the normal force $F_Z$ and the shear forces $F_X$ and $F_Y$. (c) Coils and further required electronics are embedded on flexible PCBs, resulting in a matrix setup with numerous individual sensing nodes to be manufactured in few steps. Reproduced with permission, from [42]. Copyright 2022, IEEE.

The previously discussed works have either utilized dedicated aluminum disks or the copper layer of flexible PCBs as the target material to change the inductance. However, with the metal layer positioned at the top of the insole, it can interfere with the wearer's foot, potentially causing discomfort. The inductance dependency can be integrated directly into the spacer by adding ferromagnetic particles to the elastomer, commonly referred to as magnetorheological elastomers (MREs), which provide a softer triaxial insole solution without a metal layer. The MRE insole developed by Ishiguro et al. [114] is 8 mm thick, comprising 4-mm EVA foam and 4-mm silicone rubber. The MRE is embedded in cavities within the silicone rubber, positioned above four coils per sensor, as presented in a previous work [111]. The inductance is measured and wirelessly transmitted to a PC, however, the insole is neither characterized nor tested with participants. While the MRE used by the authors contains 20% iron particles by volume, more advanced particle morphologies offer a high aspect ratio, which enables the production of anisotropic magnetic properties with locally adjusted stray fields [115].

To the best of our knowledge, there are currently no wireless multiaxial insole systems using MREs with sensor arrays. This combination can be explored in future research. Overall, inductive insoles are gaining popularity due to their relatively simple manufacturing process, incorporating multiple sensor elements, and technological advances in flexible PCB production technologies.

## D. Piezoelectric

The dynamic nature of the piezoelectric effect is frequently leveraged in gait analysis and is particularly prominent in force plates, which are mostly built on piezoelectric ceramic sensors [116]. Both hard ceramic sensors and flexible polymer-based sensors are well-established for piezoelectric signal acquisition in insoles. Tahir et al. [35] conduct a comparative study of ceramic and PVDF-based piezoelectric sensors against commonly used resistive FSR sensors. Their systematic evaluation involves insoles, each equipped with a single sensor type, tested sequentially rather than simultaneously. Chandel et al. [71] present a method to compensate for typical drift in piezoelectric ceramics using a shoe-embedded IMU, which is tested with five participants.

PVDF is the most widely used material for piezoelectric polymer sensors. In comparison to piezoceramics, PVDF-based sensors offer flexibility and conformability to the human body in wearable devices. Among the first PVDF insoles is the work by Pedotti et al. [24]. The insole is tested with one participant, which reveals suitability for heel strike detection. More recent works relying on the piezoelectric properties of PVDF are from Klimiec et al. [117], [118] and a subsequent work of the same group by Jasiewicz et al. [72] [Fig. 7(a)]. The insoles are evaluated using an electrodynamic testing setup as well as in clinical studies, including up to 20 participants. The system proves to be suitable for dynamic foot pressure assessment with clearly detectable heel strike and toe-off events. Similarly, Rajala et al. [79] utilize PVDF sensors with a charge amplifier to obtain the pressure from eight locations beneath the foot. The stationary insole is evaluated in single-step trials with five participants. In addition, they exploit four separate PVDF sheets with different dominant piezoelectric directions to calculate for plantar shear stresses [119]. Deng et al. [77] introduce a piezoelectric insole incorporating PVDF pressure sensors with a sensitivity of $23\,\mathrm{pC\,N^{-1}}$, describing it as a self-powered system. They use a 2-kg metal cylinder to simulate foot rollover, but no human experiments are conducted. In addition, simultaneous charging, measuring, and data transmission does not appear to be tested.

Chen et al. [63] build a piezoelectric insole with 24 sensor elements, consisting of PVDF film between two flexible PCBs [Fig. 7(b)]. In a previous work [120], the authors report using a force plate as a reference for walking experiments. However, their current work [63] appears to lack a reference system during the human trials. They report possible delamination of the layer compound due to localized stress concentrations and rapid changes, which may result in slippage between adjacent layers. However, according to the authors, a durability test over one month of regular wearing results in only slight deviations. Overall, the insole seems useful for COP tracing in healthy participants as well as patients. From the same group, Gao et al. [41] design a similar insole, comprising two PVDF layers on top of each other. The two layers are specifically designed to exhibit distinct dominant piezoelectric coefficients, as this difference enables the discrimination of multiple force components. The authors report sensitivities of approximately $690\,\mathrm{mV\,N^{-1}}$ in the normal direction and $160\,\mathrm{mV\,N^{-1}}$ for shear stress. However, cross correlation is not analyzed, and testing with only a single participant leaves open the possibility that the shear component reflects just a portion of the normal stress.

Zhu et al. [75] develop textile-integrated piezoelectric sensors to detect freezing of gait in patients with Parkinson's disease. Four ceramic sensors based on lead zirconate titanate (PZT) are integrated into a wearable cotton sock, which is treated as a triboelectric sensor and nanogenerator in five distinct areas. The sock suits for gait detection in three physically unimpaired participants, however, it is not tested on patients. Han et al. [81] employ two large PVDF sensors for recognizing motion and harvesting energy. Other flexible PVDF-based insoles are utilized for assessing gait parameters, and also for harvesting energy from walking [121], [122].

Most piezoelectric insoles implement either PVDF as a flexible polymer or brittle piezoceramics, such as PZT. However, recent developments have led to polymer-based electret sensors made from materials with lower environmental impact [123], [124], [125]. Ben Dali et al. [61] utilize eight polypropylene (PP) piezoelectrets to assess the gait of one participant during level walking. Luo et al. [126] characterize their PP-foam-based piezoelectrets in an electrodynamic instrument for the use in energy harvesting beneath the foot. Yabu et al. [60] use a commercial sensor (Picoleaf [127]) based on polylactic acid (PLA), in combination with FSRs on carbon insoles to estimate tri-axis GRFs in eleven participants, walking 10 m at normal speed. The piezoelectric sensor was introduced to the market in 2024 but has not yet been systematically evaluated in scholarly works.

Latsch et al. [57] embed four PLA-based piezoelectret sensors into a monolithically 3D-printed insole that is characterized using a universal testing machine and evaluated during gait with one participant on an instrumented treadmill [Fig. 7(c)]. The highly sensitive piezoelectrets are customizable via 3D printing to suit individual user and application needs. They can detect heel-strike and toe-off events faster than instrumented treadmills or commercial insoles. Furthermore, the same sensors can be used for muscle activity monitoring to detect motion intention and gait phases [125], [128].

For piezoelectric sensor insoles, it is essential to consider their dynamic behavior, since quasi-static measurements alone are insufficient because these sensors are inherently less effective for measuring static forces. Although this property constrains their application range, it also emphasizes the high sensitivity that can be achieved with piezoelectric insoles. Their sensitivity is highlighted by applications in cardiovascular monitoring, notably in insole-based ballistocardiography, where embedded piezoelectric sensors can reliably detect heartbeats and estimate respiratory rate [129], [130], [131].

### E. Triboelectric

Other less frequently used technologies for plantar pressure sensing exploit the triboelectric effect. Zhu et al. [75] combine triboelectric and piezoelectric sensing, which is described in Section III-D on piezoelectric principles. Lin et al. [73] employ a triboelectric insole with two active areas at heel and forefoot to detect heel strike and toe-off events in open-circuit mode. The insole is evaluated with one participant in several scenarios, including walking, running, and jumping. Hu et al. [56] evaluate a single participant using a triboelectric insole equipped with four sensors at each measurement point, where a PDMS dome structure and four dedicated electrodes enable the decoupling of normal and shear stresses. For an inclined loading of 30°, the measured outputs exhibit an essentially linear relationship for applied forces up to 8 N. Most other triboelectric foot wear is found in the area of energy harvesting rather than pressure sensing [132].

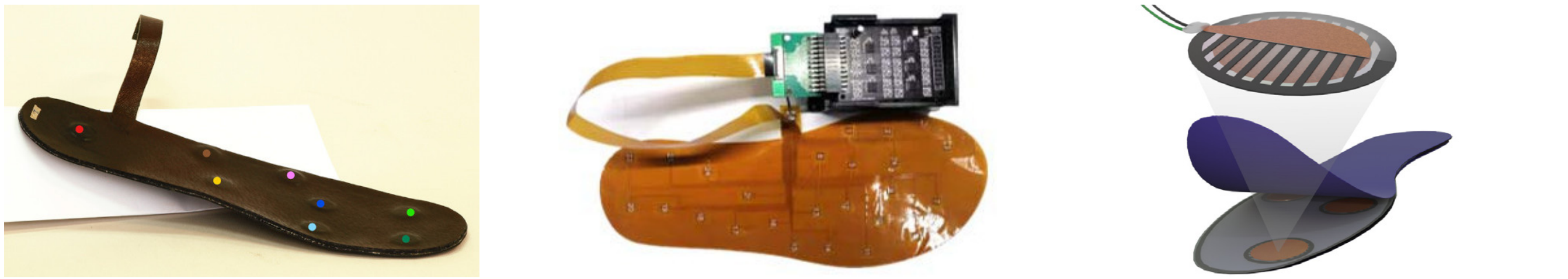

Fig. 7. Piezoelectric sensor insoles are mostly based on ceramic materials or flexible polymer films. (a) Eight PVDF sensors embedded in a leather insole for clinical application by Jasiewicz et al. [72]. Reproduced in accordance with the CC BY-NC-ND 4.0 license, from [72]. (b) Chen et al. [63] build a piezoelectric insole with 24 sensor elements, consisting of PVDF film between two flexible PCBs. Reproduced with permission, from [63]. Copyright 2021, IEEE. (c) Latsch et al. [57] employ four polymeric piezoelectret sensors monolithically 3D printed into an insole for detecting gait events. Reproduced with permission, from [57]. Copyright 2024, IEEE.

### F. Optical

Optical solutions promise benefits due to their lightweight structure and immunity against EMI. Pergolini et al. [133] use a patented technology based on optical paths that are occluded by deforming silicone rubber. The walking experiment with nine patients, diagnosed with Parkinson's disease, results in a mean absolute error of 3.36% of stance duration for the detection of heel strikes, in reference to a force plate. The detection error equals 30 ms of later detection for both heel strike and toe-off events. Since the relative timing of both events exhibits 1.62% error, the authors conclude that their insole is suitable for assessing spatiotemporal parameters in Parkinson's disease gait.

Vilarinho et al. [80] utilize optical fiber Bragg grating sensors in a cork insole to detect four steps on one participant. The COP trajectory is seen shifting from the heel toward the forefoot sensors, highlighting the sensor insole's functionality. Hao et al. [67] utilize four optical fiber Bragg grating sensors in a flexible, 3D-printed insole. Since reflection wavelengths are reduced on pressure, steps are recognized during walking, and a balance shift to one side can be detected as well. Although the cost of optical components is expected to decrease with the growing demand from the telecommunications sector, they remain relatively expensive at present. Ongoing miniaturization is expected to make future wearable solutions more viable.

## IV. Comparative Analysis and Discussion

Building on the condensed state of the art in sensor insoles, we consolidate findings across different sensor technologies regarding sensor number and placement, validation against reference systems, and emerging measurement approaches.

### A. Advantages and Limitations of Sensor Technologies

Most sensor technologies used in insoles today were already employed decades ago [25]. Despite advances in technology, materials, and computing, the fundamental challenges of capturing interaction forces in affordable wearable systems remain largely unchanged. Resistive insoles, particularly those based on FSRs, are a popular choice in the academic sector due to their low design complexity. However, capacitive, inductive, and piezoelectric insoles are becoming increasingly widespread as advancements in the underlying technologies continue.

Chen et al. [17] provide a comprehensive overview of the advantages and disadvantages of the various sensor technologies. Despite being low-cost and simple to interface, resistive sensors exhibit significant creep, hysteresis, temperature drift, and relatively high power consumption. Capacitive sensors are well-suited for low-power applications but are highly sensitive to EMI. Inductive topologies allow for triaxial sensing but are complex to interface and also sensitive to EMI. Piezoelectric sensors, due to their large dynamic range, are ideal for fast-motion applications, but are generally less effective for measuring static pressure. Their self-sufficient operation and potential for energy harvesting present advantages in mobile and wearable scenarios. Optical solutions, while largely immune to EMI, hysteresis, and creep, are not yet feasible for wearable applications, however, they may become viable in the future. Pneumatic systems, while not a primary focus of this review, also represent a promising option, particularly as advancements in 3D printing make them increasingly accessible [134].

Whether based on pneumatic [134], resistive [54], capacitive [105], or piezoelectric [57] principles, 3D-printed sensors offer a pathway to rapid and customized manufacturing [135]. Ultimately, the choice of sensor technology depends on the specific application scenario and the parameters to be measured. A promising approach for adapting sensor insoles to diverse use cases involves multimodal systems that combine two or more sensing technologies to complement their strengths.

### B. Multimodal Sensing

Sensor fusion holds promise in overcoming the limitations of individual sensor modalities. Coates et al. [136] present a multimodal wearable system integrating bioimpedance, force, temperature, and humidity sensors. Hybrid sensors for GRF measurements are also developed, combining resistive and capacitive modalities [137].

Utilizing multiple, often stacked technologies increases the overall insole thickness. In order to maintain a compact design, sensors can be distributed across the insole instead, although this may reduce spatial resolution. Piezoresponsive foam combines resistive properties for static load application and piezoelectric output for dynamic changes in one material [138]. Even though the authors motivate the application in an insole, the results are yet to be published. Other options to combining static and dynamic sensing modalities in human-related assessments involve piezoelectric-capacitive tactile sensors [139] and piezoelectric-piezoresistive nanofiber-woven fabrics [140], which may also be employed in future insole applications.

All multimodal sensor insoles included in this review incorporate piezoelectric modalities, likely reflecting the suitability of piezoelectric sensing for capturing dynamic measurements, combined with one other sensor type for static acquisition. Overall, multiple modalities can be advantageous in specific scenarios or to enhance versatility across applications.

### C. Force Measurement Capabilities

One advantage of sensor insoles over force plates is their ability to provide spatially resolved pressure data. However, GRF acquisition using insoles is inherently an estimation, as the sensors typically cover only parts of the plantar surface. Parallel load paths through the surrounding material can bypass the sensing elements, reducing accuracy.

In order to minimize bypassed load, insole designs should aim to maximize sensor coverage across the plantar area, particularly if accurate total force measurements are required. However, force estimation is derived from the sum of multiple sensor outputs, which also means that individual pressure errors accumulate, degrading the accuracy of the total GRF estimation. As a consequence, increasing spatial resolution by adding more sensors may lead to a higher cumulative error in total force estimation.

A potential solution is to optimize the insole either for spatial resolution or for total force accuracy. This design tradeoff is reflected in two commercial products by Novel. Loadsol features only up to three sensor areas and prioritizes high accuracy in total force measurements, with a sensor coverage of more than 99.6% of the plantar surface [20]. In contrast, Pedar is designed for detailed pressure distribution analysis with high spatial resolution but lower accuracy in total force estimation.

An alternative approach involves embedding load cells into shoe soles, ensuring that forces are routed entirely through the sensors. While this setup improves accuracy, it often results in bulkier footwear, which may interfere with natural gait due to increased sole thickness and weight. Despite this limitation, such sensorized shoes can serve as mobile reference systems for calibrating sensor insoles during nonstraight ambulation, e.g., turning or multidirectional walking, which is not feasible with unidirectional instrumented treadmills.

### D. Sensor Number and Placement

Although sensor resolution varies widely with the number of sensors ranging from two [64] up to 280 [40], most studies employ at least three to four sensors to obtain a minimum level of spatial detail while balancing cost and system complexity. In some scenarios, such as synchronizing an assistive device to the gait cycle, a single heel sensor may be sufficient [141]. For pathological gait analysis, additional sensor positions may be required [142]. In general, wearable devices should be designed as simply as possible to meet the needs of the application while fulfilling minimum design criteria and keeping complexity low. Higher sensor density can lead to increased crosstalk, regardless of the sensor technology, as the closer proximity of sensors and the need for multiplexed interfacing contribute to signal interference.

The most common sensor locations in insoles, starting with the most frequently used, are beneath the heel, the first metatarsal head, the fifth metatarsal head, and the big toe. This distribution reflects both frequent use in research and clinical relevance, as pressure patterns in these areas provide key insights into gait and balance. Due to their arrangement along both the anterior-posterior and medial-lateral axes of the plantar surface, these four positions are suited for basically estimating two-dimensional pressure shifts and the COP. However, this setup may not capture pressure variations in less prominent regions of the foot, such as the midfoot or lateral arch, thus, lacking positional accuracy during foot rollover. Enhancing generalizability requires a higher number of sensors, which also helps account for anatomical differences and varying gait patterns across individuals.

Ramirez-Bautista et al. [36] propose dividing the plantar surface into 15 sections, but recommend using a larger number of sensors, each smaller than 5 mm, to enable accurate peak pressure acquisition. Particularly, the extraction of gait parameters such as COP path width and length may require a minimum number of sensors [143]. By modeling the effects of sensor size and position on measured plantar pressure accuracy, Pataky [144] demonstrates that no universal spatial resolution is adequate across typical use cases. Instead, spatial resolution needs to be tailored to the specific application, taking into account the expected geometry of peak-pressure regions. Furthermore, an optimal sensor placement is not necessarily uniform but can profit from concentrated sensors in regions of high interest.

From our literature review, there is a noticeable trend correlating the number of sensor elements with the number of study participants (Table II). Generally, as the number of individual sensor elements increases, the number of participants decreases. Studies involving custom insoles developed outside of clinical settings tend to have particularly small sample sizes. We assume that custom insoles may be too complex for clinical handling, are not yet mature enough for study applications, or lack the durability required for prolonged use. As a guideline, we suggest increasing the number of participants, particularly in proof-of-concept studies with an engineering background, to ensure some variability across the population for a meaningful analysis with practical relevance. For more specific numbers, we refer the reader to Section IV-E.3.

### E. Verification and Validation Approaches

Verifying sensors involves comparing the calibrated sensor output to a known reference, whereas validation demonstrates their suitability for the intended application by testing against predefined requirements. In the context of sensor insoles, the key requirements are as follows. First, the sensor must withstand and measure the pressures exerted by full dynamic body weight, which can be significantly larger than the static body weight, particularly during activities such as running. In addition, concentrated pressures must be taken into account, occurring during specific phases of the gait cycle, such as at the heel during heel strike. Second, the sensor must accommodate the compliant, uneven plantar surface, which demands either structural flexibility or a sufficiently small sensor footprint to conform effectively. While the first requirement can be assessed through physical and technical testing, the second involves subjective human factors, directly relating to perception and comfort. Therefore, evaluation may require qualitative methods, such as user questionnaires, or human-in-the-loop approaches to assess device acceptance and perceived usability [145]. Even though approaches to quantify comfort exist in the literature [146], we focus on directly measurable quantities in this review, i.e., the physical examination.

*1) Experimental Sensor Characterization:* In order to support the physical testing, an exemplary calculation can help estimate the pressures expected beneath the foot (calculations in Supplementary Note SN1). During normal walking, an 80-kg person exerts up to 950 N of dynamic force on the heel, which corresponds to 120% of their static body weight [147]. A typical heel surface with a diameter of 5 cm spans around 20 $cm^2$, resulting in a pressure up to 480 kPa. Many insole sensors are round, with diameters ranging from 10 [56] to 20 mm [57], corresponding to an area of 0.8 to 3.1 $cm^2$, which translates to test forces across the sensor area of up to 150 N. With larger sensor sizes, the minimum test forces required for an 80-kg normally walking person exceed 150 N.

In the literature, individual sensors are characterized with smaller forces than expected during normal gait. Most works fail this requirement by orders of magnitude with test forces as small as 10 [61], 20 [58], or 28 N [56]. Some, however, test their sensors with extended pressure up to 1.2 MPa to show off their limits [94]. This pressure can even account for the impact during running, where forces exceed up to three times the body weight. According to a study that compares pressure profiles of different shoes at athletes with an average weight of more than 100 kg using the commercial Pedar insole, running can produce peak pressures of more than 500 kPa at the big toe and first metatarsal head [28]. We recommend testing and calibrating with a pressure of at least 500 kPa to cover most participants and elevated ambulation speeds. For sensors smaller than 20 mm in diameter, this pressure corresponds to a force of 150 N if evenly distributed across the contact area. Depending on the transducer principle used and the intended application, reasonable characteristics reported in the literature include sensitivity, minimal detectable change, measurement range, zero drift, full-scale drift, hysteresis, linearity, repeatability, total accuracy, and response timing [57], [63], [73], [87]. We encourage future studies to report comprehensive objective sensor properties in their characterization to facilitate comparability across works.

In addition to testing aimed at characterizing sensor properties, ISO/TS 20955 references several standards describing test methods for conventional insole footwear, including delamination resistance, abrasion resistance, water absorption and desorption, and dimensional stability. These standards are likewise applicable and useful for evaluating sensor insoles, particularly in the context of clinical adoption, where longevity and robustness are essential.

*2) Gait-Replicating Sensor Verification:* Most custom insole prototypes are not verified against standard reference systems during actual gait (Fig. 8). Instead, their characterization typically occurs prior to human testing. Human walking forces are replicated in the literature by manually applying loads with standard weights [58], [62], [64] or automatically with mechanical testing machines [57], [74], [94]. However, the quasi-static methods may not accurately reflect the dynamic force conditions encountered during real ambulation. In order to overcome this restriction, Acharya et al. [70] built a specialized pneumatic apparatus to approximate human gait dynamics. Samarentsis et al. [148] use standard weights and a motor to dynamically load their sensor, however, without a reference to record the dynamic force application.

A more generalized approach for sensor verification under controlled yet dynamic conditions employs electrodynamic testing machines, such as those manufactured by Zwick-Roell (Ulm, Germany) or Instron (Norwood, MA, USA), which are capable of applying forces exceeding 1 kN at frequencies up to 100 Hz. These systems are considered by the authors to be the most appropriate solution for simulating gait dynamics in a reproducible and human-independent manner. Nevertheless, their adoption in sensor verification remains limited, appearing only occasionally in practice [126], presumably due to the substantial cost associated with acquiring such devices. Although universal testing machines are primarily designed for quasi-static applications, they may offer a reasonable compromise [57], even when operating close to their performance limits during dynamic gait simulations.

*3) Participant-Based Validation:* While well-suited for repeated, gait-replicating characterization of individual sensors, these approaches neglect the variability of human foot soles acting on the sensor. Typically, sensor systems and the measured object exert mutual influence on each other, an effect known as measurement disturbance. In the case of sensor insoles, the human interferes with the calibrated sensor system through mechanically variable properties, undefined contact interfaces, and a complex mechanical coupling, arising from multiple loading paths between skin, soft tissue, and bone. Conversely, the sensor system can affect human behavior, both psychologically, by making the user aware of being observed, and physically, by altering the familiar contact interface.

Latsch et al. [57] demonstrate that the stiffness of the interfacing material strongly influences the sensitivity of the structured polymer sensors used. Soft interfaces that approximate human tissue increase sensitivity by up to a factor of six compared to the metallic surfaces typically used in testing machines. Foot sole stiffness and viscoelastic properties vary across individuals and, within individuals, across different phases of the gait-related loading process [149]. This finding can also be extended to prosthetics, where the critical interface is shifted from the foot sole to the socket-stump contact. These results highlight the importance of including foot sole properties in the development and validation process of sensor insoles, as this aspect appears to be largely neglected from an engineering perspective. Future work in this area should focus on investigating and optimizing both the sensor itself and the interface. Consequently, experiments involving human participants are indispensable for the proper validation of sensor insoles. Most works with a clinical background employ at least 15 participants in their studies using custom prototypes [72], [78], [88]. The normal walking speeds used are typically around $1.3\,\mathrm{m\,s^{-1}}$ [57], [147]. These numbers can serve as a reference and guidance for future studies, since many investigations from an engineering perspective include fewer participants and use arbitrary speeds, limiting the generalizability of their findings.

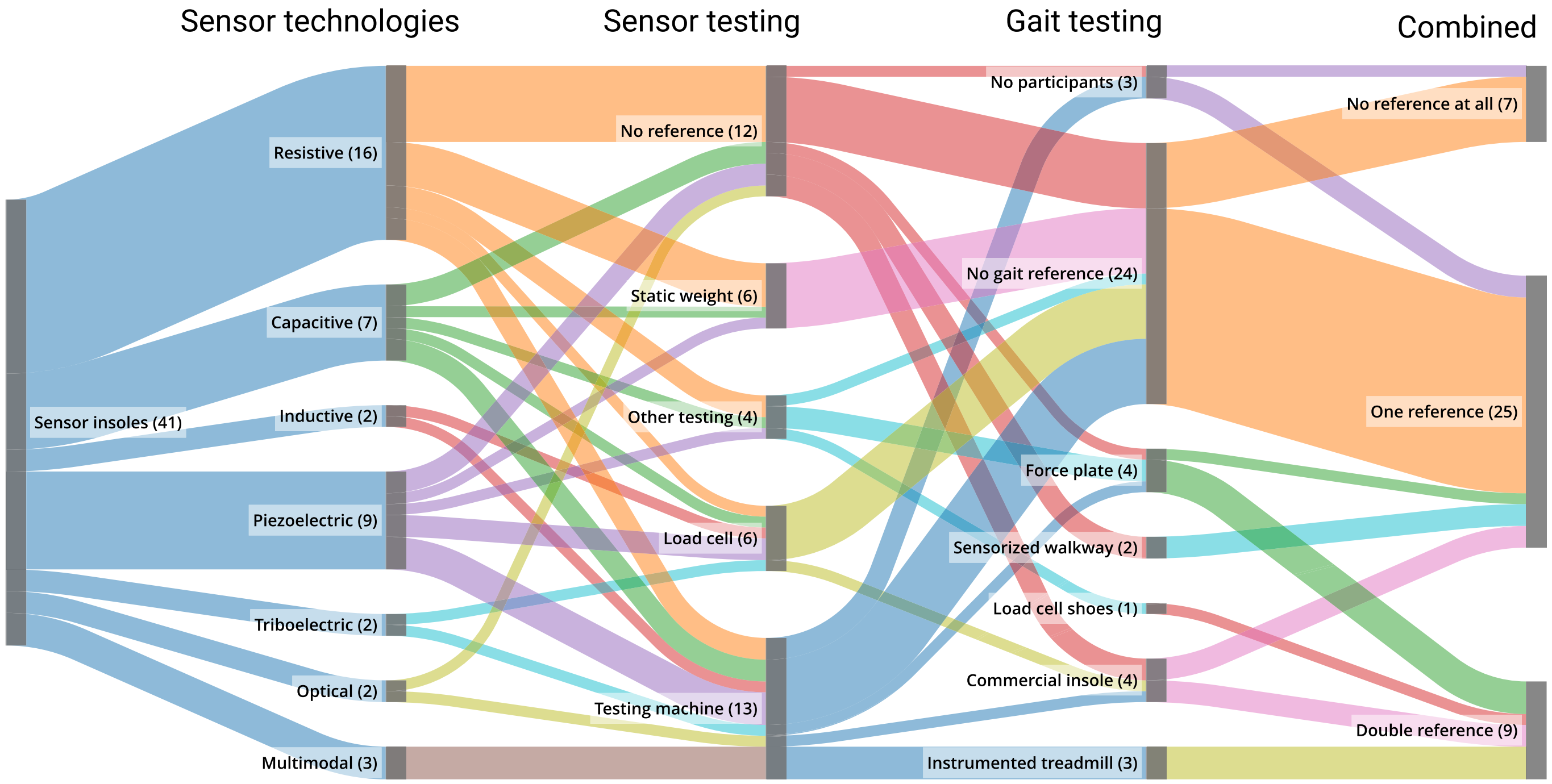

Fig. 8. Review of the reference systems utilized to calibrate and validate the 41 sensor insoles across all sensor technologies discussed. Sensor technologies are grouped by principle and prevalence, from passive to active transducers, followed by optical and multimodal. Sensor, gait, and combined testing utilities are ordered top to bottom, with a general tendency from inferior to superior reference. Only nine out of 41 works use a reference for both sensor testing and gait experiments. One reference system either for sensor testing or gait experiments is used by 25 works. Seven works are based on raw or uncalibrated sensor data as either no references are used or participant studies are missing altogether. In all three works employing instrumented treadmills as a gait reference, the sensors are characterized regarding their physical characteristics using commercial testing machines, which we propose as the gold standard for sensor insole validation. Note that, although the Sankey visualization may give that impression, not all multimodal sensor insoles are tested on instrumented treadmills (bottom row). This figure was created with the help of SankeyDiagram.net, based on the data reported in Supplementary Material Table S1.

In the literature, various reference systems are employed for gait experiments, including force plates, sensorized walkways, sensor insoles, and instrumented treadmills (Fig. 8). These systems can be classified as performing continuous or individual step measurements and as either mobile or stationary. Some studies apply their custom insoles simultaneously with commercial references, such as the F-Scan insole [33], [34], while others utilize force plates for individual step evaluation [55]. This procedure considers the dynamics of gait, which is an improvement to static-only tests, but either lacks a high-accuracy reference in the case of commercial insoles or continuous measurements in the case of force plates, which both simultaneously would be highly preferred from the perspective of measurement validation. Consequently, instrumented treadmills, with their highly accurate integrated force plates and ability for continuous gait experiments, represent the best option when the protocol permits stationary laboratory setups. Otherwise, commercial insoles are the most practical alternative, although their mobile deployment comes at the cost of reduced accuracy.

When using force plates as a reference, whether floor-mounted or integrated into treadmills, it is essential to align the coordinate systems of the mobile insole and the stationary plate. Due to the insole's flexibility, even the coordinate system of individual sensor elements shifts relative to others during foot rollover. This misalignment is particularly relevant when comparing or calibrating three-dimensional force components, because the vertical GRF recorded by an insole sensor is not perpendicular to the floor throughout the gait cycle, for example, during touch-down and lift-off phases.

In conclusion, the most advanced prototype evaluation begins with fundamental sensor characterizations using a testing machine, ideally including dynamic load application, followed by participant trials using the custom sensor insoles on an instrumented treadmill as a reference. Three out of 41 works in this review follow these recommendations [54], [57], [60]. Six studies employ a combination of two other reference systems, resulting in a total of nine works with double reference (Fig. 8, detailed in Supplementary Table S1). A single reference system, used either for sensor testing or for gait experiments, is reported in 25 of the reviewed works. Among these, only five employ a reference system during gait experiments; the remaining 20 include a loading reference for sensor testing, with 18 omitting a reference during their gait experiments and two conducting no gait experiments at all. Seven of the reviewed works make no use of reference systems or lack gait testing altogether. In the broader literature, the actual number of such omissions is expected to be considerably higher, given that the works reviewed here are preselected based on the inclusion of human studies or other distinctive characteristics. We encourage authors of future studies to incorporate both machine-driven dynamic loading for objective characterization of sensor properties and reference-backed gait experiments involving a sufficiently large participant cohort. Only this combination can ensure reproducible sensor performance across human variability to support progression toward clinical adoption.

### F. Multiaxial Sensors

The growing interest in shear-force measurements in clinical and research contexts has led to the development of multiaxial film-based pressure sensors [119], [150]. Piezoelectric [41], triboelectric [56], and inductive [42] sensor insoles appear capable of measuring shear forces in addition to normal forces. Since the influence of shear pressure distribution beneath the foot remains insufficiently investigated, assessing both perpendicular components with triaxial sensor matrices can provide a more complete picture. Further research in this area is likely to be informed by advances in related fields such as tactile sensing, where shear force measurements are also critical, albeit under significantly lower load conditions than those encountered during ambulation [110], [111].

Although further research on this topic is promising, validating multiaxial sensors remains challenging, as it requires extended testing setups capable of measuring shear forces. Even in rigid integrated transducers, decoupling multiple force components is complex and continues to be the subject of current research [151], [152], [153]. When using mechanically flexible polymers, the main challenge lies in minimizing directional crosstalk, while achieving high sensitivity in the primary direction and maintaining overall flexibility. In addition, insole designs must account for the asymmetric force distribution, i.e., large vertical body weight versus small horizontal shear components. One potential approach is to embed sufficiently small rigid transducers within a conformable matrix. Other strategies aim to compensate for directional crosstalk through analytical methods or machine learning [154].

Although related patents have been filed [103], no commercial multiaxial sensor insoles appear to be available. Therefore, future insole concepts should consider extending measurements to shear forces, while ensuring proper characterization and validation. In the absence of a mobile multiaxial reference system, we suggest first verifying the sensor response under dynamic loading using a stationary multiaxial testing machine, followed by gait experiments for validation, with either a mobile normal-force reference or stationary multiaxial equipment such as an instrumented treadmill.

## V. Conclusion

This review highlights the current landscape of sensor insole technologies with a focus on their physical and functional characteristics, rather than downstream applications such as machine learning or classification tasks. While many studies use insoles for data-driven analysis, they often lack rigorous sensor verification and validation, which is essential for applications that require deterministic control, such as assistive devices. We propose a standard based on testing machines and instrumented treadmills to ensure comparability across studies.

Compared to stationary force plates, sensor insoles offer key advantages that open new application fields. They enable pressure distribution mapping and support out-of-lab gait analysis in natural environments, extending beyond just straight and level walking patterns. However, they are typically less accurate in measuring total GRFs, for which force plates remain better suited.

Despite rapid advancements in material science and sensor design, clinical studies incorporating noncommercial, state-of-the-art insoles remain scarce. This gap may be attributed to the differing priorities between research domains, where engineers aim for innovation, clinicians often favor simplicity and reliability to ensure practical benefit for patients. Thorough validation of insole prototypes is required to effectively apply novel sensor technologies to clinical and real-world use. Testing capabilities must include not only static conditions but more importantly dynamic loading of at least 500 kPa that reflects natural gait patterns. Furthermore, the uncertainty introduced by varying foot tissue stiffness should be considered in future designs.

The proposed standard for sensor insole and study design is valid for a multitude of application areas. As a guideline, the literature indicates that 15 participants and sensor insoles with 15 sensor elements of 5-mm size provide a reasonable compromise between complexity and generalizability. The ideal sensor insole combines the ability to measure both normal and shear forces, maintains a thin and flexible profile to preserve natural gait, requires minimal external power, and is validated for clinical use. However, such a multipurpose insole may not be the optimal solution for each individual scenario, where specialized designs better meet specific requirements while minimizing complexity. As development in this field continues, the insights provided in this review aim to guide future efforts toward building more reliable, accurate, and application-specific systems.

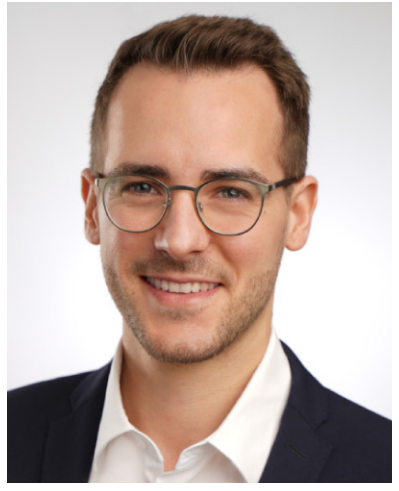

**Bastian Latsch** (Member, IEEE) received the M.Sc. and Ph.D. degrees in electrical engineering from Technische Universität Darmstadt, Darmstadt, Germany, in 2019 and 2024, respectively.

Since then, he has been a Postdoctoral Researcher with the Measurement and Sensor Technology Group, Technische Universität Darmstadt. His research interests include sensors in human movement assistance and interfaces for human–machine interaction.

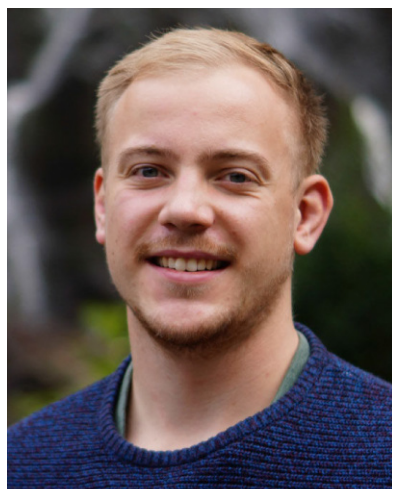

**Felix Herbst** received the M.Sc. degree in electrical engineering from Technische Universität Darmstadt, Darmstadt, Germany, in 2022, where he is currently pursuing the Ph.D. degree in electrical engineering with the Measurement and Sensor Technology Group.

His research interests include the design of structurally integrated mechanical sensors and robotics in medical applications.

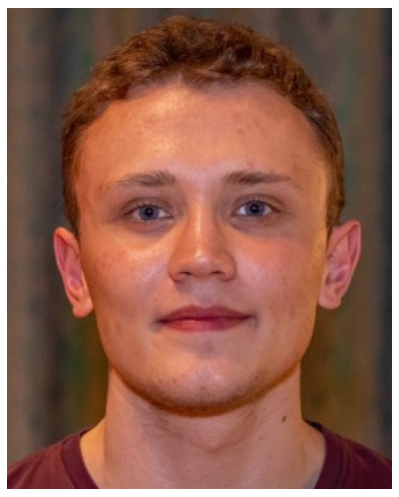

**Mark Suppelt** (Graduate Student Member, IEEE) received the M.Sc. degree in electrical engineering from Technische Universität Darmstadt, Darmstadt, Germany, in 2024, where he is currently pursuing the Ph.D. degree in electrical engineering with the Measurement and Sensor Technology Group.

His research interests include embedded systems integrating electronics, sensors, and software, along with assistive devices supporting human locomotion.

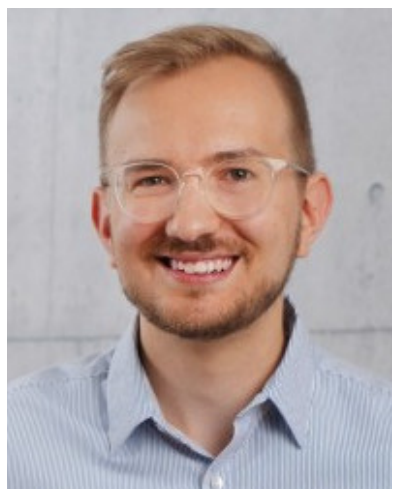

**Julian Seiler** (Graduate Student Member, IEEE) received the M.Sc. degree in electrical engineering from Technische Universität Darmstadt, Darmstadt, Germany, in 2023, where he is currently pursuing the Ph.D. degree in electrical engineering with the Measurement and Sensor Technology Group.

His research interests include assistive devices supporting human locomotion, vibrotactile feedback, and human–robot interaction.

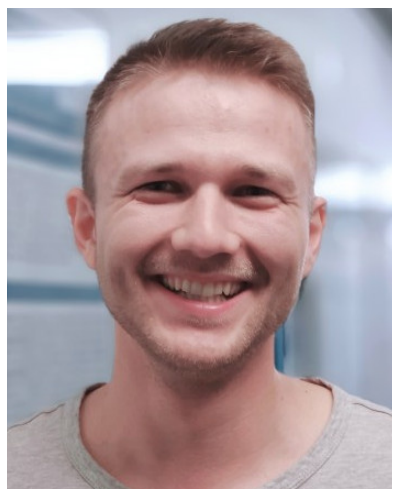

**Stephan Schaumann** (Graduate Student Member, IEEE) received the M.Sc. degree in mechatronics from Technische Universität Darmstadt, Darmstadt, Germany, in 2022, where he is currently pursuing the Ph.D. degree in electrical engineering with the Measurement and Sensor Technology Group.

His research interests include sensors in human movement assistance, ferroelectret-based transducers, and ultrasonic transducers.

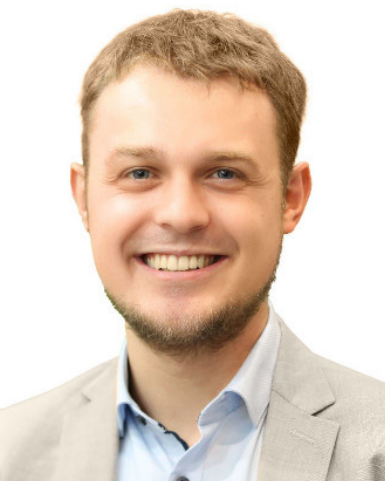

**Sven Suppelt** (Graduate Student Member, IEEE) received the M.Sc. degree in electrical engineering from Technische Universität Darmstadt, Darmstadt, Germany, in 2021, where he is currently pursuing the Ph.D. degree in electrical engineering with the Measurement and Sensor Technology Group.

His research interests include electronics, firmware, and software for embedded systems, with a strong emphasis on miniaturizing and integrating force sensors.

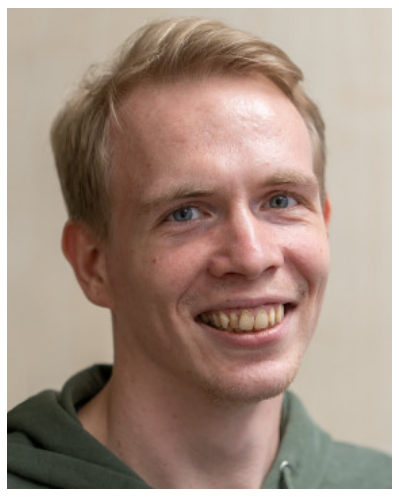

**Alexander A. Altmann** (Graduate Student Member, IEEE) received the M.Sc. degree in information systems technology from Technische Universität Darmstadt, Darmstadt, Germany, in 2021.

Since 2022, he has been a Research Associate with the Measurement and Sensor Technology Group, Technische Universität Darmstadt. His research interests include the additive manufacturing of flexible sensors based on ferroelectrets and the integration into human-centric applications, such as textile integration or medical applications.

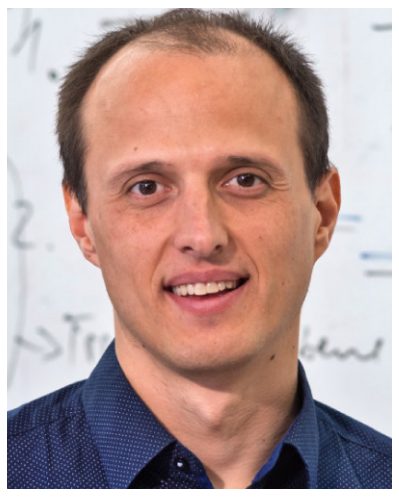

**Martin Grimmer** received the Diploma degree in sports science from FSU, Jena, Germany, in 2008, and the Ph.D. degree from Technische Universität Darmstadt, Darmstadt, Germany, in 2015.

He was a Guest Researcher with Arizona State University, Tempe, AZ, USA, in 2012, and a Postdoctoral Researcher with Harvard University, Cambridge, MA, USA, in 2015, and with ETH Zürich, Zürich, Switzerland, in 2016. He is currently an Athene Young Investigator Group Leader with Technische Universität Darmstadt. His research interests include biomechanics of human movement, the transfer of biological-inspired mechanical principles to wearable robotics, human–robot interaction, and wearable robotics control.

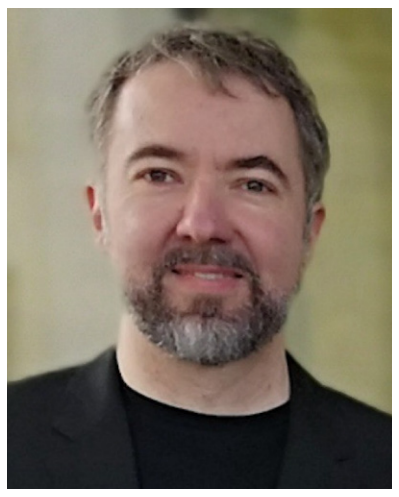

**Mario Kupnik** (Senior Member, IEEE) received the Diplom-Ingenieur degree in electrical engineering from Graz University of Technology, Graz, Austria, in 2000, and the Ph.D. degree in electrical engineering from the University of Leoben, Leoben, Austria, in 2004.

From 2005 to 2011, he was a Postdoctoral Researcher, a Research Associate, and a Senior Research Scientist at Edward L. Ginzton Laboratory, Stanford University, Stanford, CA, USA. From 2011 to 2014, he was a Full Professor of electrical engineering at Brandenburg University of Technology, Cottbus, Germany. Since 2015, he has been a Full Professor with Technische Universität Darmstadt, Darmstadt, Germany, where he is currently the Head of the Measurement and Sensor Technology Group. His research topics are micromachined sensors and actuators, multiphysics simulations, ferroelectrets, flowmetering of gases and liquids, ultrasound, electroacoustics, human–machine interaction, and robotics.